\documentclass[12pt,a4paper]{article}
\usepackage{epsfig}
\usepackage{amsmath}
\usepackage{amsfonts}
\usepackage{amssymb}
\usepackage{graphicx}
\usepackage{slashed}
\usepackage{color}
\usepackage{textcomp}

\definecolor{blu}{cmyk}{1,0.7,0,0.6}

\begin{document}

\begin{titlepage}
\noindent \flushright{August 2015} \vspace{0.2cm}
\begin{center}
\begin{Large}
\begin{bf}
{\color{blu}Scalar Sector Phenomenology of Three-Loop Radiative
Neutrino Mass Models}\\
\end{bf}
\end{Large}
\end{center}
\vspace{0.2cm}
\begin{center}
\begin{bf}
{Amine~Ahriche,$^{a,b,}$\footnote{aahriche@ictp.it} Kristian~L.
McDonald$^{c,}$\footnote{klmcd@physics.usyd.edu.au} and
Salah~Nasri$^{d,}$\footnote{snasri@uaeu.ac.ae}}\\
\end{bf}
\vspace{0.5cm}
\begin{it}
$^a$ Department of Physics, University of Jijel, PB 98 Ouled Aissa, DZ-18000 Jijel, Algeria\\
\vspace{0.1cm}
$^b$ The Abdus Salam International Centre for Theoretical Physics, Strada Costiera 11, I-34014, Trieste, Italy\\
\vspace{0.1cm}
$^c$ ARC Centre of Excellence for Particle Physics at the Terascale,\\
School of Physics, The University of Sydney, NSW 2006, Australia\\
\vspace{0.1cm} $^d$ Physics Department, UAE University, POB 17551,
Al Ain, United Arab Emirates \vspace{0.3cm}
\end{it}
\vspace{1.5cm}
\end{center}
\begin{abstract}
We perform a phenomenological study of the scalar sector of two
models that generate neutrino mass at the three-loop level and
contain viable dark matter candidates. Both models contain a charged
singlet scalar and a larger scalar multiplet (triplet or
quintuplet). We investigate the effect of the extra scalars on the
Higgs mass and analyze the modifications to the triple Higgs
coupling. The new scalars can give observable changes to the Higgs
decay channel $h\rightarrow\gamma \gamma$ and, furthermore, we find
that the electroweak phase transition becomes strongly first-order
in large regions of parameter space.\\ \vspace{1cm} \textbf{PACS}:
12.60.Fr, 12.60.-I, 95.35.+d.
\end{abstract}
\vspace{1cm}
\end{titlepage}

\section{Introduction\label{sec:introduction}}

The Standard Model (SM) of particle physics is a spectacularly successful
theory that stands as one of the truly great scientific achievements.
Despite this success, however, the theory possesses a number of
short-comings, suggesting it will likely require extensions and/or
modifications in the future. The most obvious motivation for extending the
SM is the need to incorporate gravity. However, a lack of present-day
experimental guidance makes the pursuit of the theory of quantum gravity an
incredibly challenging task.

Additional evidence that the SM is incomplete comes from the experimental
observation of neutrino mixing and the need to explain the missing
gravitating (i.e.~dark) matter, required on galactic scales. These puzzles
motivate the addition of new particle species to the SM and there is much
hope that such new species will manifest in future experiments (in
particular at Run II of the LHC). The neutrino mass and dark matter (DM)
problems have stimulated much research and there are multiple candidate
solutions one can pursue. The problems may have independent solutions,
though it seems reasonable to ask whether the two puzzles could share a
unified or common solution. Could the neutrino mass and DM problems be
related?

In 2002, Krauss, Nasri and Trodden (KNT) proposed a simple extension of the
SM model that admits a relationship between the existence of DM and the
origin of neutrino mass~\cite{Krauss:2002px}. In this approach, one adds new
fields to the SM, such that neutrino mass is generated radiatively at the
three-loop level, with one of the particles propagating in the mass diagram
being a DM candidate. The model employs two charged singlet scalars, $%
S^{+}_{1,2}$, and three generations of gauge-singlet fermions $N$. A $Z_2$
symmetry with action $\{S_2^+,\, N\}\rightarrow \{-S_2^+,\,- N\}$ is also
imposed. This ensures stability of the lightest fermion $N$, thereby giving
a DM candidate, and also prevents a coupling between SM neutrinos and $N$,
which would otherwise generate tree-level neutrino masses. The result is a
type of unified description for the origin of neutrino mass and DM, with the
removal of the DM candidate simultaneously turning off neutrino mass.%
\footnote{%
For recent studies of the KNT model see Refs.~\cite%
{Baltz:2002we,Cheung:2004xm,Ahriche:2013zwa,Ahriche:2014xra}.}

In recent years, a number of basic generalizations of the KNT model have
appeared. In one such model (hereafter `the triplet model'), the singlet
fields $S_2^+$ and $N$ are replaced by $SU(2)_L$ triplets~\cite%
{Ahriche:2014cda}. This model retains the $Z_2$ symmetry to ensure DM
stability and prevent tree-level neutrino masses, and gives a viable
alternative unified framework for the DM and neutrino mass problems. A
further generalization exchanges the singlet fields $S_2^+$ and $N$ for $%
SU(2)_L$ quintuplet fields~\cite{Ahriche:2014oda}. This model (hereafter
`the quintuplet model') does not require the $Z_2$ symmetry in order to
prevent tree-level neutrino masses, and is a viable theory for radiative
neutrino mass, independent of DM considerations. Interestingly, the
most-general Lagrangian for the model contains a single $Z_2$-breaking
coupling $\lambda$. When taken small, $\lambda\ll1$, the model gives a
long-lived DM candidate, while turning $\lambda$ off completely, $%
\lambda\rightarrow0$, activates a $Z_2$ symmetry and gives absolutely stable
DM. Thus, the quintuplet model is a viable model of radiative neutrino mass,
with or without DM.

Due to the presence of larger multiplets with non-trivial $SU(2)$ charges in
both the triplet and quintuplet models, the phenomenology of the models is
rather rich. In the present work, we extend the analysis of Refs.~\cite%
{Ahriche:2014cda,Ahriche:2014oda} and undertake a more extensive study of
the phenomenology of both models. We investigate the effect of the triplet
and quintuplet scalars on both the Higgs mass and the triple Higgs coupling,
showing that the latter can experience sizable modifications. We also study
the effect of the new multiplets on the Higgs decay channels $h\rightarrow
\gamma \gamma ,\gamma Z$. Our work shows that, e.g.~observable changes are
expected to $\mathcal{B}(h\rightarrow \gamma \gamma )$, with some regions of
parameter space already excluded for the triplet model. The effect of the
enlarged scalar sector on the electroweak phase transition is also analyzed,
revealing a tendency for a strongly first-order phase transition in large
regions of parameter space.

Before proceeding we note that further generalizations of the KNT model are
possible. Ref.~\cite{Chen:2014ska} presented colored generalizations and
other related three-loop models that employ slightly modified loop
topologies. A septuplet generalization of the KNT model was proposed in Ref.~%
\cite{Ahriche:2015wha}. This had the interesting feature of automatically
containing an absolutely stable DM candidate, without requiring a new
symmetry. A minimal scale-invariant implementation was also recently studied~%
\cite{Ahriche:2015loa}. More generally, a number of authors have studied
connections between radiative neutrino mass and DM in recent years, see
e.g.~Refs.~\cite{Ma:2006km}-\cite{Baek:2015mna}.

This paper is structured as follows. In Section~\ref{sec:models}, we outline
the triplet and quintuplet models, describing some key features and
elucidating some stability constraints on the scalar potentials. We study
the influence of the new multiplets on the Higgs mass and the triple Higgs
coupling in Section~\ref{sec:higgs_mass_and_coupling}. The electroweak phase
transition is considered in Section~\ref{sec:phase_transition} and we turn
to the Higgs decay channels $h\rightarrow \gamma\gamma$ and $h\rightarrow
\gamma Z$ in Section~\ref{sec:higgs_decays}. Conclusions are presented in
Section~\ref{sec:conc}.

\section{Three-Loop Radiative Neutrino Masses \label{sec:models}}

The SM employs the gauge symmetry $\mathcal{G}_{SM}=SU(3)_{c}\times
SU(2)_{L}\times U(1)_{Y}$. In this work, we consider extensions of the SM
that include the charged singlet scalar $S\sim (1,1,2)$, the scalar
multiplet $T\sim (1,2n+1,2)$ and three generations of chiral beyond-SM
fermions, $\mathcal{F}_{i}\sim (1,2n+1,0)$, where $i=1,2,3,$ labels
generations and numbers in parenthesis denote charges under $\mathcal{G}%
_{SM} $. We use the integer $n=0,1,2,$ to label the distinct models. The
case with $n=0$ is the KNT model, for which all beyond-SM fields are $%
SU(2)_{L}$ singlets: $S\equiv S_{1}^{+}$, $T\equiv S_{2}^{+}$ and $\mathcal{F%
}\equiv N$. For $n=1$ ($n=2$) the multiplets $T$ and $\mathcal{F}$ are $%
SU(2)_{L}$ triplets (quintuplets) and one has the triplet (quintuplet)
model. In all cases, the new multiplets are subject to a discrete symmetry
with action $\{T,\,\mathcal{F}\}\rightarrow \{-T,\,-\mathcal{F}\}$. This
ensures a stable DM candidate, which should be taken as the lightest fermion
$\mathcal{F}_{1}^{0}\equiv \mathcal{F}_{\text{{\tiny DM}}}$.\footnote{%
For $n=0$ the scalar $T$ is charged, while for $n=1$ and $n=2$ the neutral
component of the scalar $T$ has non-trivial couplings to the $Z$ boson and
id therefore excluded as a DM candidate by direct-detection constraints.}
Detailed analysis of the DM annihilation channels appears in Refs.~\cite%
{Ahriche:2014cda,Ahriche:2014oda}.

With the aforementioned particle content, the Lagrangian contains the
following terms:
\begin{equation}
\mathcal{L}\supset\{f_{\alpha\beta}\,\overline{L_{\alpha}^{c}}%
\,L_{\beta}\,S^{+}+g_{i\alpha}\,\overline {\mathcal{F}_{i}}\,T\,e_{\alpha R}+%
\mathrm{H.c}\}\;-\;\frac{1}{2}\,\overline{\mathcal{F}_{i}^{c}}\,\mathcal{M}%
_{ij}\,\mathcal{F}_{j}\;-\;V(H,S,T). \label{eq:lagrangian}
\end{equation}
Here, $L_{\alpha}\sim(1,2,-1)$ are SM lepton doublets, $e_{\alpha
R}\sim(1,1,-2)$ are the SM charged lepton singlets and $f_{\alpha\beta}=-f_{%
\beta\alpha}$ denote Yukawa couplings. Lepton flavors are labeled by
lower-case Greek letters, $\alpha, \beta \in\{e,\,\mu,\,\tau \}$. The
singlet-leptons $e_{\alpha R}$ couple to the exotics $T$ and $\mathcal{F}$
through the Yukawa matrix $g_{i\alpha}$, and the superscript
\textquotedblleft$c$" is used to denote charge conjugation. The Lagrangian
shows that both $T$ and $\mathcal{F}$ are sequestered from SM neutrinos. We
denote the SM Higgs doublet as $H\sim(1,2,1)$.

\begin{figure}[t]
\begin{center}
\includegraphics[width = 0.55\textwidth]{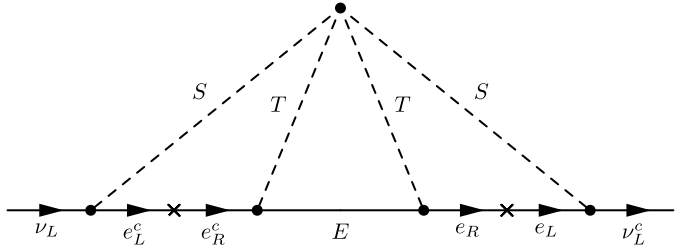}
\end{center}
\caption{Three-loop diagram for neutrino mass. Here, $S\sim (1,1,2)$ and $%
T\sim (1,2n+1,2)$ are beyond-SM scalars while $\mathcal{F}\sim (1,2n+1,0)$
is a beyond-SM fermion. The case with $n=0$ corresponds to the KNT model~%
\cite{Krauss:2002px}, while $n=1$ gives the triplet model~\cite{Ahriche:2014cda} and $n=2$ gives the quintuplet model~%
\cite{Ahriche:2014oda}. } \label{fig:nu_mass}
\end{figure}

The combination of the Yukawa terms in Eq.~\eqref{eq:lagrangian} and the
term $\sim S^{2}(T^{\ast })^{2}$ in the scalar potential $V(H,\,S,\,T)$
(discussed below) explicitly breaks lepton-number symmetry. The models
therefore generate radiative neutrino masses, which appear at the three-loop
level as shown in Figure~\ref{fig:nu_mass}. Due to the $Z_{2}$ symmetry, the
neutral components of the exotic fermions $\mathcal{F}$ do not mix with SM
neutrinos at any order in perturbation theory, and similarly there is no
mixing between charged leptons and $\mathcal{F}$. In both the triplet and
quintuplet models, the charged scalar $S$ can be within reach of TeV scale
collider experiments~\cite{Ahriche:2014cda,Ahriche:2014oda}.

\subsection{Triplet Model}

The case with $n=1$ gives the triplet model, for which $\mathcal{F}_{i}\sim
(1,3,0)$ and $T\sim (1,3,2)$ are $SU(2)_{L}$ triplets. We write the triplet
fields in symmetric-matrix notation as $T=T_{ab}$ and $\mathcal{F}=\mathcal{F%
}_{ab}$, where $a,b\in \{1,\,2\}$ are $SU(2)_{L}$ indices. The multiplets
contain the following components \cite{Ahriche:2014cda}
\begin{align}
& \ \mathcal{F}_{11}=\mathcal{F}_{L}^{+},\quad \mathcal{F}_{12}=\mathcal{F}%
_{21}=\frac{1}{\sqrt{2}}\mathcal{F}_{L}^{0},\quad \mathcal{F}_{22}=\mathcal{F%
}_{L}^{-}\equiv (\mathcal{F}_{R}^{+})^{c}, \notag \\
& \ T_{11}=T^{++},\quad T_{12}=T_{21}=\frac{1}{\sqrt{2}}T^{+},\quad
T_{22}=T^{0},
\end{align}%
while the triplet mass term gives
\begin{equation}
-(\overline{\mathcal{F}_{i}^{c}})_{ab}\,\mathcal{M}_{ij}\,(\mathcal{F}%
_{j})_{cd}\,\epsilon ^{ac}\,\epsilon ^{bd}=-\overline{\mathcal{F}_{iR}^{+}}\,%
\mathcal{M}_{ij}\,\mathcal{F}_{jL}^{+}-\frac{1}{2}\,\overline{(\mathcal{F}%
_{iL}^{0})^{c}}\,\mathcal{M}_{ij}\,\mathcal{F}_{jL}^{0}.
\label{eq:mass_lagrangeE}
\end{equation}%
The neutral-fermion mass terms are brought to the correct sign by defining
the Majorana fermions $\mathcal{F}_{i}^{0}=\mathcal{F}_{i,L}^{0}-(\mathcal{F}%
_{i,L}^{0})^{c}$. Radiative corrections from SM gauge bosons lift the
degeneracy between the components of $\mathcal{F}$, leaving $\mathcal{F}^{0}$
as the lightest component~\cite{Cirelli:2009uv}, though for most purposes
this small splitting can be neglected~\cite{Ahriche:2014cda}. We work in the
diagonal basis with $\mathcal{M}_{ij}=\mathrm{diag}(M_{1},M_{2},M_{3})$,
where $M_{1}\equiv M_{\text{{\tiny DM}}}$ is the lightest triplet-fermion
mass. According to the analysis in Ref.~\cite{Ahriche:2014cda}, the DM mass
should lie in the range $2.35-2.75$ \textrm{TeV}, and all triplet members
masses should be larger than the DM mass. Accordingly, the triplet scalars
are beyond the reach of the LHC, though, as we shall see, they can still be
probed indirectly.

The scalar potential for the triplet model has the form
\begin{align}
& V(H,S,T)=-\mu ^{2}|H|^{2}+\lambda |H|^{4}+\mu _{S}^{2}\ |S|^{2}+\frac{%
\lambda _{S}}{2}|S|^{4}+\mu _{T}^{2}[(T^{\ast })^{ab}T_{ab}]+\frac{\eta _{1}%
}{2}[(T^{\ast })^{ab}T_{ab}]^{2} \notag \\
& +\frac{\eta _{2}}{2}(T^{\ast })^{ab}T_{bc}(T^{\ast })^{cd}T_{da}+\lambda
_{SH}|S|^{2}|H|^{2}+\left\{ \bar{\lambda}_{ST}|S|^{2}+\bar{\lambda}%
_{HT}|H|^{2}\right\} [(T^{\ast })^{ab}T_{ab}] \notag \\
& -\lambda _{HT}(H^{\ast })^{a}T_{ab}(T^{\ast
})^{bc}H_{c}+\frac{\lambda _{ST}}{4}(S^{-})^{2}T_{ab}T_{cd}\epsilon
^{ac}\epsilon ^{bd}+\frac{\lambda _{ST}^{\ast
}}{4}(S^{+})^{2}(T^{\ast })^{ab}(T^{\ast })^{cd}\epsilon
_{ac}\epsilon _{bd}. \label{V3}
\end{align}%
Vacuum stability requires that the quantities
\begin{equation}
\lambda ,~\lambda _{S},~\eta _{1}+\eta _{2},~\eta _{1}+\tfrac{1}{2}\eta
_{2},~\left\vert
\begin{array}{cc}
\lambda & \left( \lambda _{HT}-\bar{\lambda}_{HT}\right) ^{0} \\
\left( \lambda _{HT}-\bar{\lambda}_{HT}\right) ^{0} & \eta _{1}+\eta _{2}%
\end{array}%
\right\vert ,~\left\vert
\begin{array}{cc}
\lambda _{S} & \bar{\lambda}_{ST}^{0} \\
\bar{\lambda}_{ST}^{0} & \eta _{1}+\tfrac{1}{2}\eta _{2}%
\end{array}%
\right\vert , \label{c3}
\end{equation}%
are taken strictly positive, with $\lambda _{\#}^{0}=\min (\lambda _{\#},0)$%
. Taking the charged scalar squared mass $\mu _{S}^{2}$ and the scalar
triplet squared masses as positive ensures the absence of spontaneous charge
symmetry breaking and guarantees that $\left\langle T^{0}\right\rangle =0$.
The latter is is necessary to preserve the $Z_{2}$ symmetry and retain a
stable DM candidate.

\subsection{Quintuplet Model}

The quintuplet model corresponds to $n=2$, in which case one has $\mathcal{F}%
_i\sim(1,5,0)$ and $T\sim(1,5, 2)$ as $SU(2)_L$ quintuplets. In
symmetric-matrix notation, the quintuplets are written as $T_{abcd}$ and $%
\mathcal{F}_{abcd}$, where~\cite{Ahriche:2014oda}
\begin{align}
& \ \mathcal{F}_{1111}=\mathcal{F}_{L}^{++},\ \mathcal{F}_{1112}=\frac{%
\mathcal{F}_{L}^{+}}{\sqrt{4}},\ \mathcal{F}_{1122}=\frac {\mathcal{F}%
_{L}^{0}}{\sqrt{6}},\ \mathcal{F}_{1222}=\frac{(\mathcal{F}_{R}^{+})^{c}}{%
\sqrt{4}},\ \mathcal{F}_{2222}=(\mathcal{F}_{R}^{++})^{c}, \notag \\
& \ T_{1111}=T^{+++},\ T_{1112}=\frac{T^{++}}{\sqrt{4}},\ T_{1122}=\frac{%
T^{+}}{\sqrt{6}},\ T_{1222}=\frac{T^{0}}{\sqrt{4}},\ T_{2222}=T^{-}.
\end{align}
Observe that $T^{+}$ and $T^{-}$ are distinct fields with $%
T^{-}\neq(T^{+})^{\ast}$. The explicit expansion of the fermion mass term
gives
\begin{equation}
-\frac{1}{2}\,(\overline{\mathcal{F}_{i}^{c}})_{abcd}\,\mathcal{M}_{ij}\,(%
\mathcal{F}_{j})_{efgh}\,\epsilon^{ae}\,\epsilon^{bf}\,\epsilon
^{cg}\,\epsilon^{dh}+\mathrm{H.c.}=-\overline{\mathcal{F}_{i}^{++}}\,%
\mathcal{M}_{ij}\,\mathcal{F}_{j}^{++}-\overline{\mathcal{F}_{i}^{+}}\,%
\mathcal{M}_{ij}\,\mathcal{F}_{j}^{+}-\frac{1}{2}\overline{\mathcal{F}%
_{i}^{0}}\,\mathcal{M}_{ij}\,\mathcal{F}_{j}^{0},
\end{equation}
where $\mathcal{F}^{0}$ is a Majorana fermion and the other four components
of $\mathcal{F}$ combine to give two charged (Dirac) fermions:
\begin{equation}
\mathcal{F}^{++}=\mathcal{F}_{L}^{++}+\mathcal{F}_{R}^{++}\,,\quad \mathcal{F%
}^{+}=\mathcal{F}_{L}^{+}-\mathcal{F}_{R}^{+}\,,\quad \mathcal{F}^{0}=%
\mathcal{F}_{L}^{0}+(\mathcal{F}_{L}^{0})^{c}.
\end{equation}%
Without loss of generality, we again employ the basis with $\mathcal{M}_{ij}=%
\mathrm{diag}(M_{1},\,M_{2},\,M_{3})$, where $M_{1}\equiv M_\text{{\tiny DM}}
$. According to the analysis of Ref.~\cite{Ahriche:2014oda}, the DM mass is
expected to lie in the range $5.65-6.95$~\textrm{TeV}, and all quintuplet
members should have masses that exceed $M_\text{{\tiny DM}}$.\footnote{%
The degeneracy between neutral and charged components of $\mathcal{F}$ is
again lifted by radiative corrections.} Consequently the quintuplet scalars
cannot be produced directly at the LHC.

The full scalar potential for the quintuplet model that respects the global
symmetry $Z_{2}$ is given by
\begin{align}
V(H,\,S,\,T)& =-\mu ^{2}|H|^{2}+\lambda |H|^{4}+\mu _{S}^{2}\ |S|^{2}+\frac{%
\lambda _{S}}{2}|S|^{4}+\lambda _{SH}|S|^{2}|H|^{2}+\mu _{T}^{2}[(T^{\ast
})^{abcd}T_{abcd}] \notag \\
& +\frac{\eta _{1}}{2}[(T^{\ast })^{abcd}T_{abcd}]^{2}+\frac{\eta _{2}}{2}%
[(T^{\ast })^{abcd}T_{cdef}(T^{\ast })^{eflm}T_{lmab}]+\frac{\eta _{3}}{2}%
[(T^{\ast })^{abcd}T_{bcde}(T^{\ast })^{eflm}T_{aflm}] \notag \\
+& \left\{ \lambda _{ST}|S|^{2}+\lambda _{HT1}|H|^{2}\right\} [(T^{\ast
})^{abcd}T_{abcd}]+\lambda _{HT2}(T^{\ast })^{abcd}T_{ebcd}(H^{\ast
})^{e}H_{a} \notag \\
& +\frac{\kappa }{4}(S^{-})^{2}T_{abcd}T_{efgh}\epsilon
^{ae}\epsilon ^{bf}\epsilon ^{cg}\epsilon ^{dh}+\mathrm{H.c.}
\label{V5}
\end{align}%
Vacuum stability requires that the following quantities%
\begin{gather}
\lambda ,~\lambda _{S},~\eta _{1}+\tfrac{1}{2}\eta _{2}+\tfrac{1}{2}\eta
_{3},~\eta _{1}+\tfrac{1}{2}\eta _{2}+\tfrac{5}{8}\eta _{3},~\eta _{1}+\eta
_{2}+\eta _{3}>0, \notag \\
\left\vert
\begin{array}{cc}
\lambda & \left( \lambda _{HT1}+\tfrac{3}{4}\lambda _{HT2}\right) ^{0} \\
\left( \lambda _{HT1}+\tfrac{3}{4}\lambda _{HT2}\right) ^{0} & \eta _{1}+%
\tfrac{1}{2}\eta _{2}+\tfrac{5}{8}\eta _{3}%
\end{array}%
\right\vert ,~\left\vert
\begin{array}{ccc}
\lambda _{S} & \lambda _{ST}^{0} & \lambda _{ST}^{0} \\
\lambda _{ST}^{0} & \eta _{1}+\tfrac{1}{2}\eta _{2}+\tfrac{1}{2}\eta _{3} &
\left( \eta _{1}+\tfrac{1}{3}\eta _{2}+\tfrac{1}{2}\eta _{3}\right) ^{0} \\
\lambda _{ST}^{0} & \left( \eta _{1}+\tfrac{1}{3}\eta _{2}+\tfrac{1}{2}\eta
_{3}\right) ^{0} & \eta _{1}+\eta _{2}+\eta _{3}%
\end{array}%
\right\vert , \label{c5}
\end{gather}%
be strictly positive. Similar to the triplet case, spontaneous charge
symmetry breaking is avoided and the neutral quintuplet remains VEV-less by
taking the squared masses of the charged scalar and the quintuplet to be
positive. This preserves the $Z_{2}$ symmetry.

In the numerical scans performed below, we impose the above mentioned
conditions such as vacuum stability, charge non-breaking and $\left\langle
T^{0}\right\rangle =0$, and also require the Higgs mass to be within the
range reported by ATLAS and CMS, $m_{h}=125.09\mp 0.21~\mathrm{GeV}$ \cite%
{Aad:2015zhl}. We restrict our attention to the perturbativity domain,
demanding that the physical vertices in Eqs.~(\ref{V3}) and (\ref{V5}) be
less than 3. For both the triplet~\cite{Ahriche:2014cda} and quintuplet~\cite%
{Ahriche:2014oda} models, we consider the charged singlet scalar mass
between 100 \textrm{GeV} and 1 \textrm{TeV}. The masses for the scalar
multiplet members should be larger than the dark matter mass, i.e., $%
M_{T}>2.35$~\textrm{TeV} and $M_{T}>5.65$~\textrm{TeV}, for the triplet and
quintuplet models, respectively. For the numerical analysis we consider
20,000 sets of benchmark points for both the triplet and quintuplet models.
The benchmarks reproduce the observed DM relic density while also achieving
viable neutrino masses and avoiding lepton flavor violating constraints (see
Refs.~\cite{Ahriche:2014cda,Ahriche:2014oda} for discussion on constraints).

\section{Higgs Mass and Triple Higgs Coupling\label%
{sec:higgs_mass_and_coupling}}

In order to estimate the Higgs mass and the triple Higgs coupling at
one-loop, it is necessary to properly define the effective potential, with
the Higgs mass being its second derivative and the triple Higgs coupling
given by the third derivative:
\begin{align}
& \left. m_{h}^{2}=\frac{\partial ^{2}}{\partial h^{2}}V_{eff}^{T=0}(h)%
\right\vert _{h=\upsilon }, \notag \\
& \left. \lambda _{hhh}=\frac{\partial ^{3}}{\partial h^{3}}%
V_{eff}^{T=0}(h)\right\vert _{h=\upsilon }.
\end{align}%
Here $h$\ is the real part of the neutral component in the doublet, $%
\upsilon $ is its VEV, and $V_{eff}^{T=0}(h)$\ is the zero temperature
one-loop Higgs effective potential. In this work we employ the $\overline{DR}%
^{\prime }$ scheme, for which the effective potential is given by~\cite{DRp}
\begin{equation}
V_{eff}^{T=0}(h)=-\frac{\mu ^{2}}{2}h^{2}+\frac{\lambda }{4}%
h^{4}+\sum_{i}n_{i}\frac{m_{i}^{4}(h)}{64\pi ^{2}}\left( \log \frac{%
m_{i}^{2}(h)}{\Lambda ^{2}}-\frac{3}{2}\right) , \label{Veff}
\end{equation}%
where $n_{i}$\ is the field multiplicity and $\Lambda $ is the
renormalization scale, which we take as the measured value of the Higgs
mass, $\Lambda =125.09~\mathrm{GeV}$ \cite{Aad:2015zhl}. The quantities $%
m_{i}^{2}(h)$ are the field-dependent squared masses (presented in the
appendix for both triplet and quintuplet models). In this class of models, $h
$ is the only scalar with a non-zero VEV, so all field-dependent masses can
be written as $m_{i}^{2}\left( h\right) =\mu _{i}^{2}+\alpha _{i}^{2}h^{2}/2$%
, for constant $\alpha _{i}$.

At tree-level, the parameter $\mu ^{2}$ in the potential is given by $\mu
^{2}=\lambda \upsilon ^{2}$. After including one-loop corrections, the
parameter $\mu ^{2}$ is corrected as%
\begin{equation}
\mu ^{2}=\lambda \upsilon ^{2}+\frac{1}{32\pi ^{2}}\sum_{i}\left.
n_{i}\alpha _{i}m_{i}^{2}\left( \ln \frac{m_{i}^{2}}{\Lambda ^{2}}-1\right)
\right\vert _{h=\upsilon ,\mu ^{2}\equiv \mu ^{2}+\delta \mu ^{2}},
\label{mu}
\end{equation}%
in order to ensure the one-loop VEV value remains as $\upsilon =246~\mathrm{%
GeV}$. The term $\delta \mu ^{2}$ represents the radiative corrections to
the $\mu ^{2}$-term, due to all fields, and is expected to be dominated by
contributions from the new heavy fields. The Higgs mass at one-loop can be
similarly defined by%
\begin{equation}
m_{h}^{2}=2\lambda \upsilon ^{2}+\frac{\upsilon ^{2}}{32\pi ^{2}}%
\sum_{i}n_{i}\alpha _{i}^{2}\log \frac{m_{i}^{2}}{\Lambda ^{2}},
\label{mh}
\end{equation}%
where the radiative corrections (i.e., second term in (\ref{mh}))\ are also
expected to be dominated by contributions from heavy new fields. Although $%
m_{h}^{2}$ is determined by experimental observations, the doublet quartic
coupling $\lambda $\ can still be (very) small, relative to the SM value,
while reproducing the observed Higgs mass. According to the size and sign of
the one-loop contribution in (\ref{mh}), the quartic coupling $\lambda $
must be smaller (larger) than the tree-level value $3m_{h}^{2}/\upsilon ^{2}$%
, for a positive (negative) loop contribution.

The triple Higgs coupling is the third derivative of (\ref{Veff}), which can
be simplified as%
\begin{equation}
\lambda _{hhh}=6\lambda \upsilon +\frac{\upsilon }{32\pi ^{2}}%
\sum_{i}n_{i}\alpha _{i}^{2}\left( \frac{\alpha _{i}\upsilon ^{2}}{m_{i}^{2}}%
+3\log \frac{m_{i}^{2}}{\Lambda ^{2}}\right) . \label{lhhh1}
\end{equation}%
Using (\ref{mu}) and (\ref{mh}), the triple Higgs coupling (\ref{lhhh1}) can
be simplified as
\begin{equation}
\lambda _{hhh}=\frac{3m_{h}^{2}}{\upsilon }\left( 1+\frac{\upsilon ^{4}}{%
96\pi ^{2}m_{h}^{2}}\sum_{i}\frac{n_{i}\alpha _{i}^{3}}{m_{i}^{2}}\right) .
\label{lhhh}
\end{equation}

\begin{figure}[t]
\begin{center}
\includegraphics[width=0.5\textwidth]{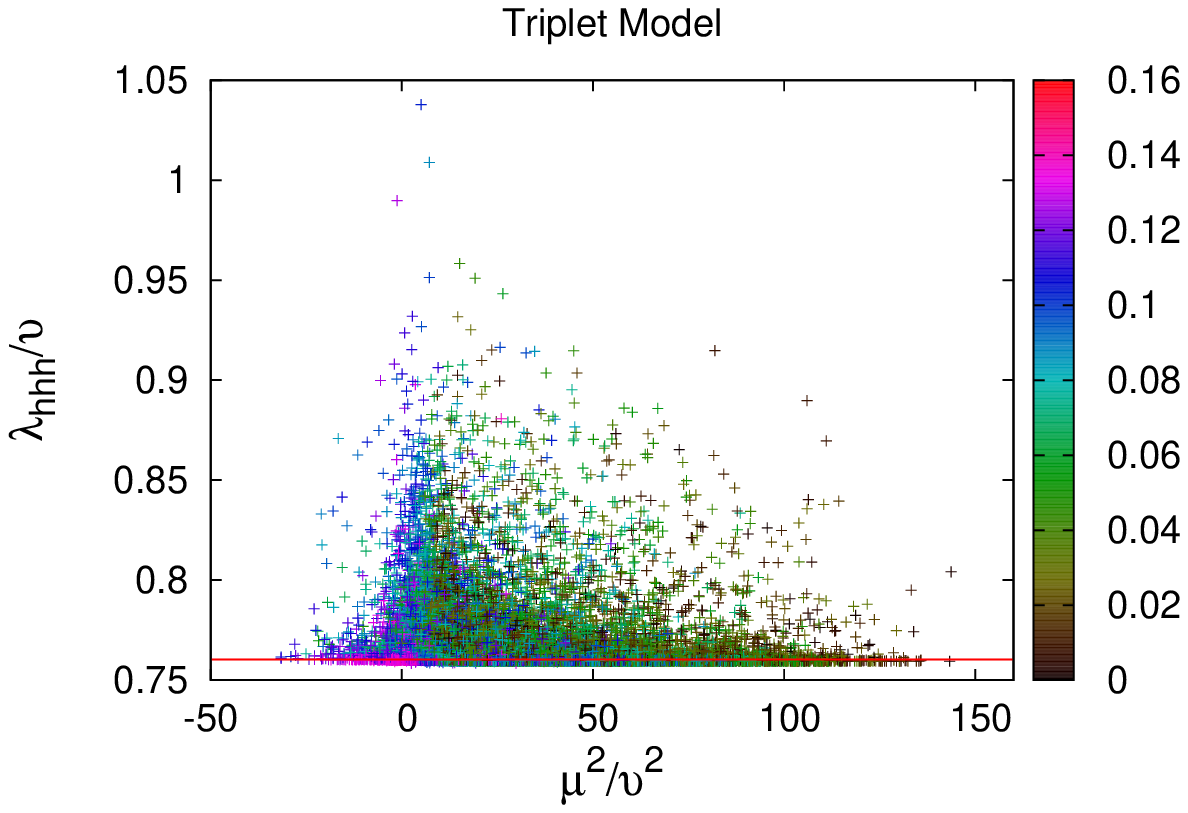}~\includegraphics[width=0.5%
\textwidth]{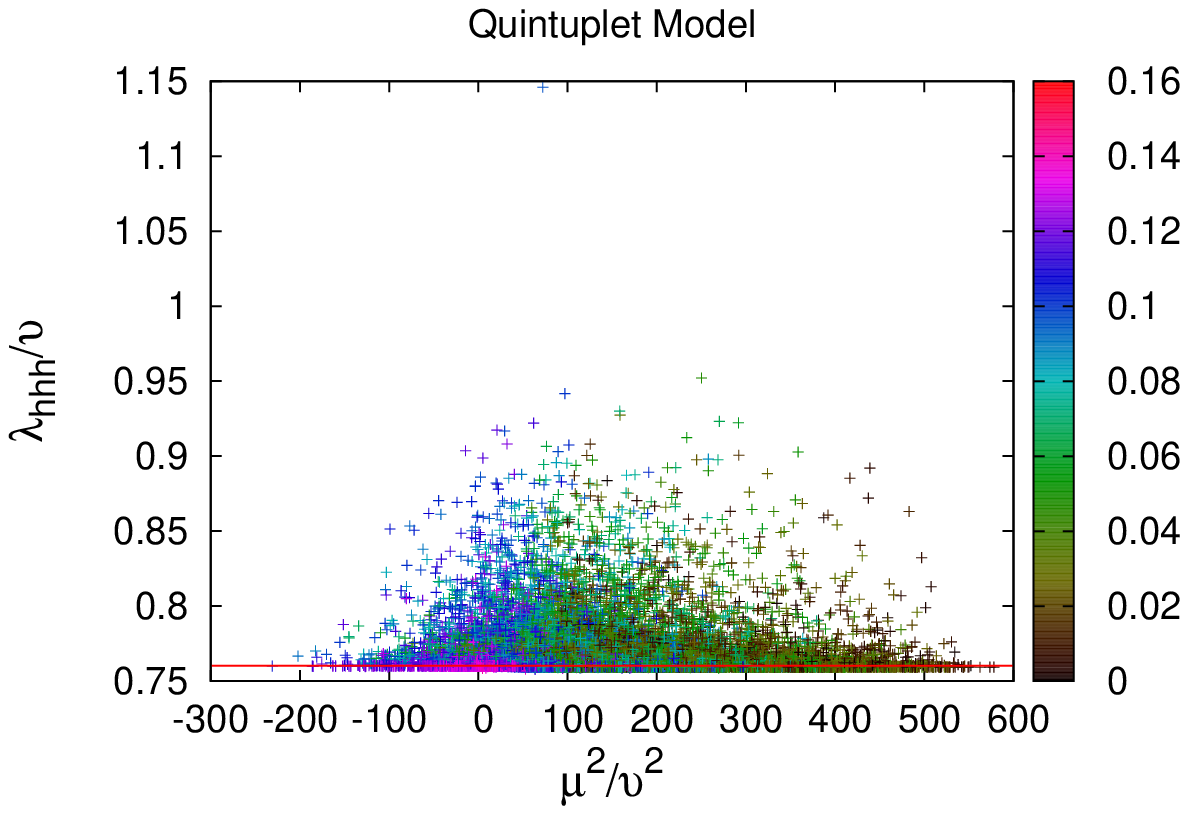} \includegraphics[width=0.5\textwidth]{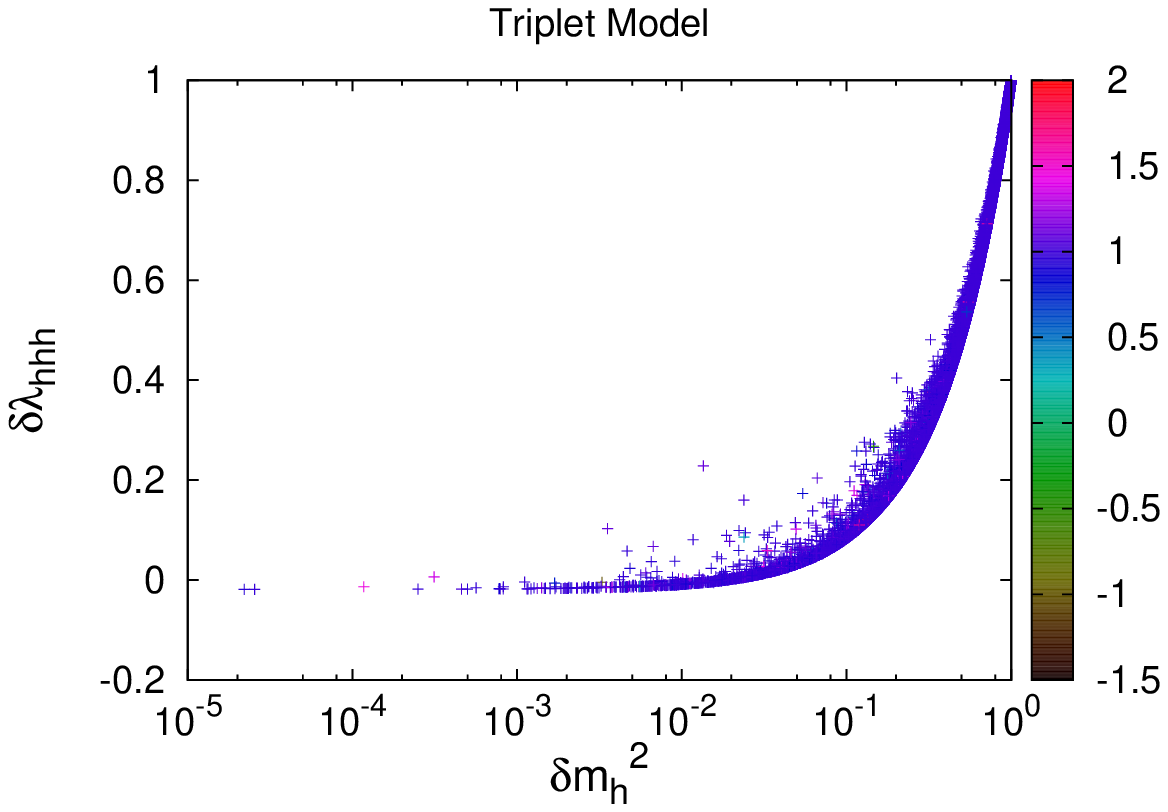}~%
\includegraphics[width=0.5\textwidth]{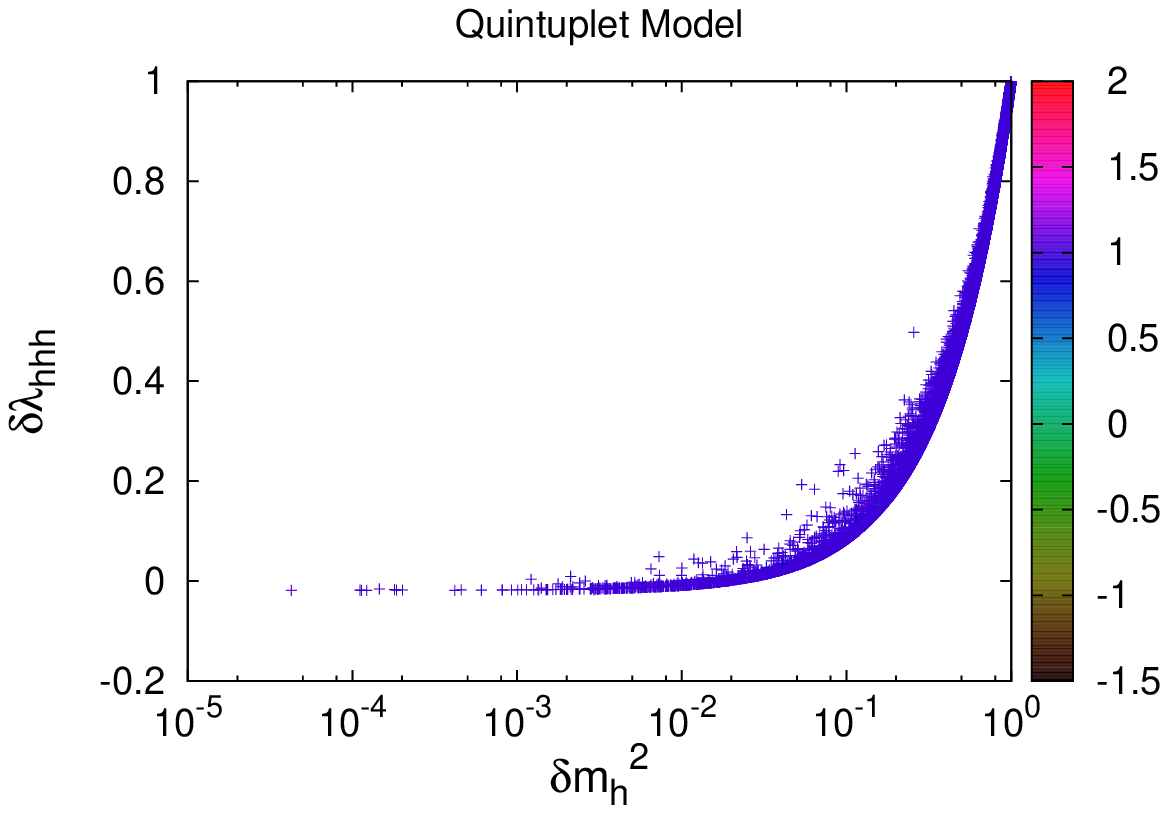}
\end{center}
\caption{\textit{Top: the triple Higgs coupling versus the mass
parameter parameter $\mu ^{2}$, both in units of the Higgs VEV, for
the triplet (left) and quintuplet (right) models. The palette gives
the Higgs quartic coupling $\lambda $, and the red line shows the SM
triple Higgs coupling value. Bottom: the relative radiative
contribution to the triple Higgs coupling versus the relative
radiative contribution to the Higgs mass. The palette gives the
relative radiative contribution to the Higgs mass parameter $\delta
\mu ^{2}$.}} \label{mmu}
\end{figure}

The size of the radiative effects can be parameterized by the following
dimensionless quantities:
\begin{equation}
\delta \mu ^{2}=\frac{\mu ^{2}-\lambda \upsilon ^{2}}{\mu ^{2}},\quad \delta
m_{h}^{2}=\frac{m_{h}^{2}-2\lambda \upsilon ^{2}}{m_{h}^{2}},\quad \delta
\lambda _{hhh}=\frac{\lambda _{hhh}-6\lambda \upsilon }{\lambda _{hhh}}.
\label{rel}
\end{equation}%
These measure the relative strength of the radiative contributions to the
Higgs bare mass-squared, the physical mass-squared $\mu ^{2}$, and the
triple Higgs coupling, respectively. Using the previously mentioned
benchmark points, in Fig.~\ref{mmu} we plot the triple Higgs coupling versus
the mass-squared parameter $\mu ^{2}$, for both triplet and quintuplet
models. We also show the relative strength of the radiative contributions to
the Higgs mass, triple Higgs coupling and the parameter $\mu ^{2}$, as
defined in Eq.~(\ref{rel}).

One notices from Fig.~\ref{mmu}-Top that the mass-squared parameter $\mu
^{2} $ can be large, even up to 100 (500) times the Higgs VEV-squared $%
\upsilon ^{2}$ for the triplet (quintuplet) models. The larger values are
required in order to balance the radiatively induced mass term in the
Lagrangian, i.e., the second term on the left-hand side of (\ref{mu}). The
radiative corrections can also be negative, depending on the value of the
Higgs quartic coupling; i.e., for $\lambda \gtrsim 0.08$. Due to the fact
that the extra fields in the quintuplet model are much heavier than those of
the triplet model, their radiative contributions are larger and therefore
the $\mu ^{2}$ parameter values are larger, as it is evident from the
figures. From Fig.~\ref{mmu}-Bottom, one notices that the relative
radiative-contributions to the Higgs mass and triple coupling are
proportional, i.e.~when the Higgs mass is completely generated radiatively,
the triple Higgs coupling is also dominated by radiative effects. We also
observe that for most of the benchmark points, in both the triplet and
quintuplet models, the mass-squared parameter $\mu ^{2}$ is fully radiative,
as shown in the palette.

\begin{figure}[t]
\begin{center}
\includegraphics[width=0.5\textwidth]{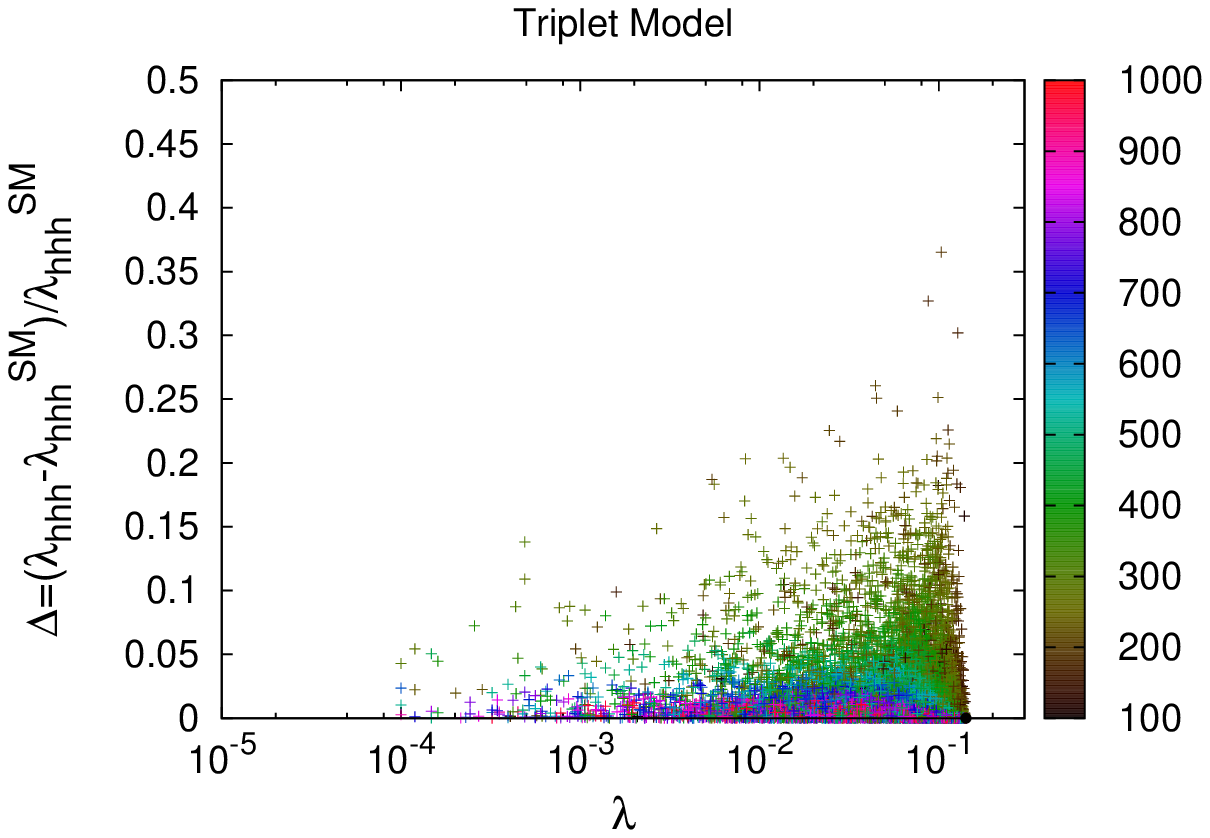}~\includegraphics[width=0.5%
\textwidth]{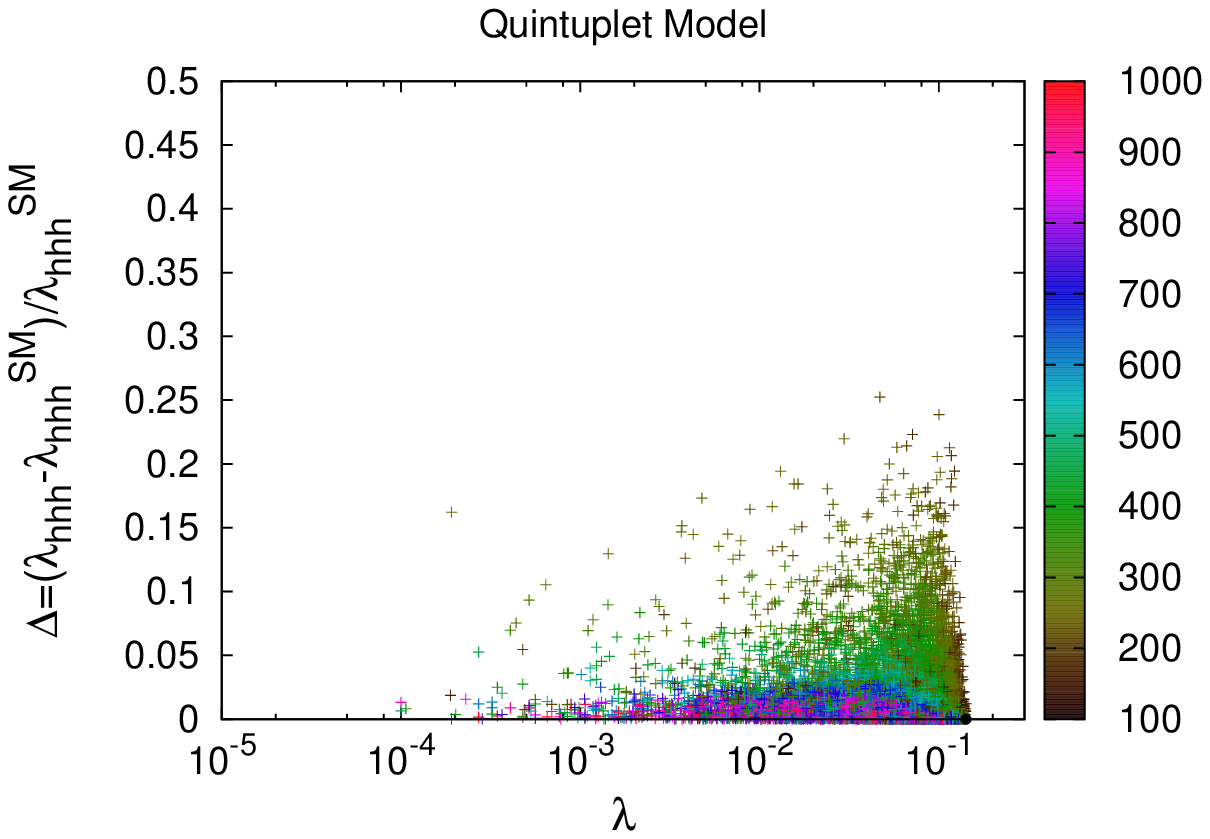}
\end{center}
\caption{\textit{The relative enhancement in the triple Higgs
coupling with respect to the SM, $\Delta $, versus the Higgs quartic
coupling. The palette shows the mass of the charged scalar $S^{+}$
in \textrm{GeV}. The black point at ($\lambda =\lambda ^{SM},0$)
refers to the SM.}} \label{DD}
\end{figure}

The relevant quantity for collider phenomenology is the relative enhancement
in the triple Higgs coupling, with respect to the SM value, which is defined
as%
\begin{equation}
\Delta =\frac{\lambda _{hhh}-\lambda _{hhh}^{SM}}{\lambda _{hhh}^{SM}}.
\end{equation}%
According to Eq.~(\ref{lhhh}), the relative enhancement of the triple Higgs
coupling is given by%
\begin{equation}
\Delta =\frac{\sum_{i\neq SM}\frac{n_{i}\alpha _{i}^{3}}{m_{i}^{2}}}{\frac{%
96\pi ^{2}m_{h}^{2}}{\upsilon ^{4}}+\sum_{i=all}\frac{n_{i}\alpha _{i}^{3}}{%
m_{i}^{2}}}. \label{D}
\end{equation}%
In Fig.~\ref{DD}, we show the relative enhancement of the triple Higgs
coupling, Eq.~(\ref{D}), for the benchmark sets used previously. The figure
shows that the relative enhancement of the triple Higgs coupling, with
respect to the SM, are larger for large values of the quartic coupling $%
\lambda $, and smaller for small values of the charged scalar mass. Also,
one notices that the relative enhancement in the triple Higgs coupling, with
respect to the SM value, is always positive, contrary to other models \cite%
{Ahriche:2015mea, Ahriche:2013vqa}, and furthermore it can exceed 35\% for
both the triplet and quintuplet models.

\section{Electroweak Phase Transition\label{sec:phase_transition}}

The SM cannot successfully explain baryogenesis \cite{EWB} for two reasons:
(1) the CP violating source in the CKM matrix is too small and (2) the
electroweak phase transition (EWPT) is not strongly first order. The latter
is required to suppress the B+L violating processes in the broken phase,
inside the bubble, when its wall is expanding during the transition. In the
SM, the criterion for a strongly first-order EWPT \cite{SFOPT},
\begin{equation}
\upsilon (T_{c})/T_{c}>1, \label{v/t}
\end{equation}%
is not fulfilled since the ratio is given by $\upsilon _{c}/T_{c}\sim
\lambda $, which would require a Higgs mass below 42 GeV~\cite{mhbound}.
Here $T_{c}$ is the critical temperature at which the effective potential
exhibits two degenerate minima, one at zero and the other at $\upsilon
(T_{c})$. Both $T_{c}$ and $\upsilon (T_{c})$ are determined using the full
effective potential at finite temperature, which is given by \cite{Th}
\begin{align}
V_{eff}(h,T)& =V_{eff}^{T=0}(h)+\tfrac{T^{4}}{2\pi ^{2}}\sum_{i}n_{i}J_{B,F}%
\left( m_{i}^{2}/T^{2}\right) +V_{ring}(h,T); \label{VT} \\
J_{B,F}\left( \alpha \right) & =\int_{0}^{\infty }x^{2}\log (1\mp \exp (-%
\sqrt{x^{2}+\alpha })){.} \label{JBF}
\end{align}%
In the above, we include an important leading term from the higher-order
loop corrections, which can play an important role during the EWPT dynamics,
namely the so-called daisy contributions \cite{daisy}%
\begin{equation}
V_{ring}(h,T)=-\frac{T}{12\pi }\sum\limits_{i}n_{i}\left\{ \tilde{m}%
_{i}^{3}(h,T)-m_{i}^{3}(h)\right\} . \label{daist}
\end{equation}%
The summation is over scalar and longitudinal gauge degrees of freedom, with
$\tilde{m}_{i}^{2}(h,T)=m_{i}^{2}(h)+\Pi (T)$ their thermal masses, and $\Pi
(T)$ are the thermal parts of the self energy (given in the appendix). For
our analysis we include the daisy contributions by following an alternative
approach to Eq.~(\ref{daist}), i.e.~by replacing the field dependent masses
of the scalar and longitudinal gauge fields by their thermal masses $\tilde{m%
}_{i}^{2}(h,T)$ in the full effective potential~(\ref{VT}). In order to
account for all the (heavy and light) degrees of freedom, we evaluate the
integrals (\ref{JBF}) numerically.

The strength of the EWPT can be improved when new bosonic degrees of freedom
are present, as occurs in the present models. It is clear from (\ref{mh})
that for large values for the couplings \{$\lambda _{SH},\lambda _{HT},\bar{%
\lambda}_{HT}$ for triplet model and $\lambda _{SH},\lambda _{HT1,2}$ for
quintuplet model\} and/or small mass-values for the extra (singlet and
multiplet) scalars, the one-loop correction to the Higgs mass can be
significant, allowing the Higgs self-coupling to be smaller. Consequently
one can fulfill the criterion (\ref{v/t}) without conflicting with recent
Higgs mass measurements \cite{Aad:2015zhl}.

Analyses of similar models~\cite{Ahriche:2013zwa, Ahriche:2015mea, hna} has
shown that extra scalars can help to generate a strongly first order EWPT
by: (a) relaxing the Higgs self-coupling $\lambda $ to be as small as $%
\mathcal{O}(10^{-4})$; and (b) enhancing the value of the effective
potential at the wrong vacuum at the critical temperature, without
suppressing the ratio $\upsilon (T_{c})/T_{c}$, which relaxes the severe
bound on the mass of the SM Higgs.

\begin{figure}[h]
\begin{center}
\includegraphics[width=0.5\textwidth]{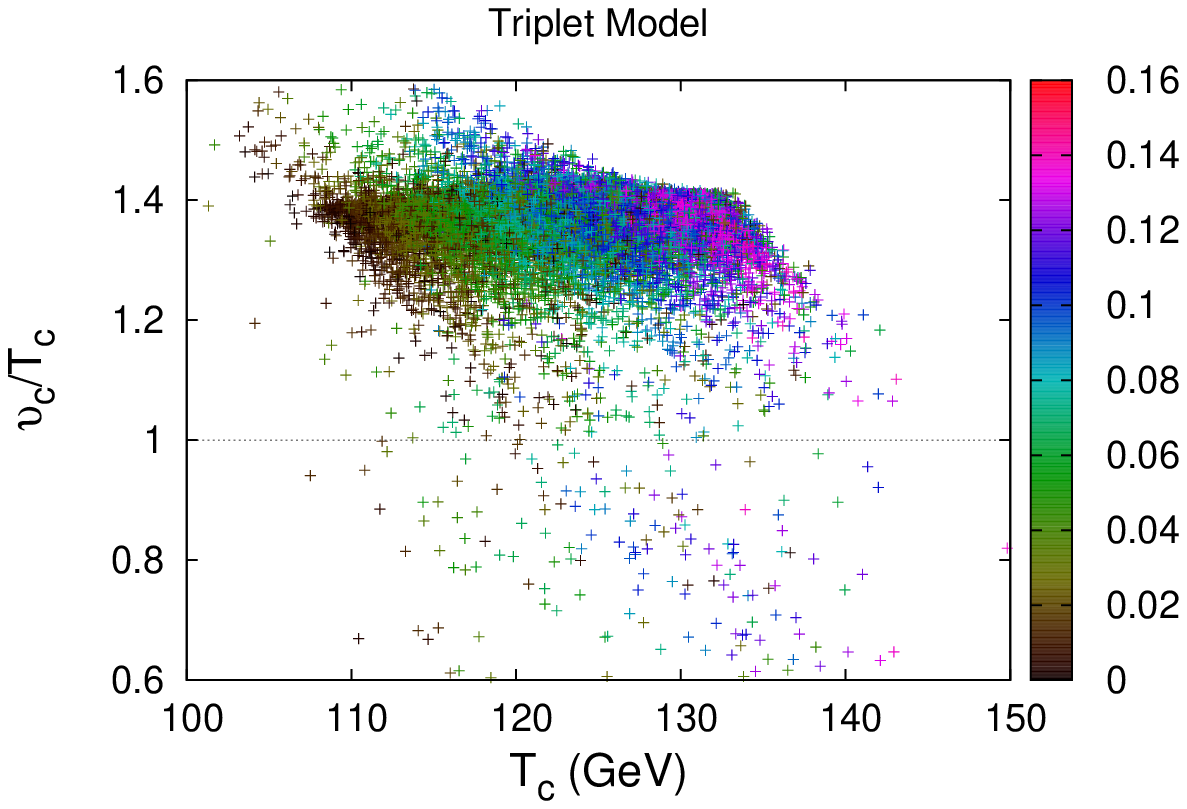}~\includegraphics[width=0.5%
\textwidth]{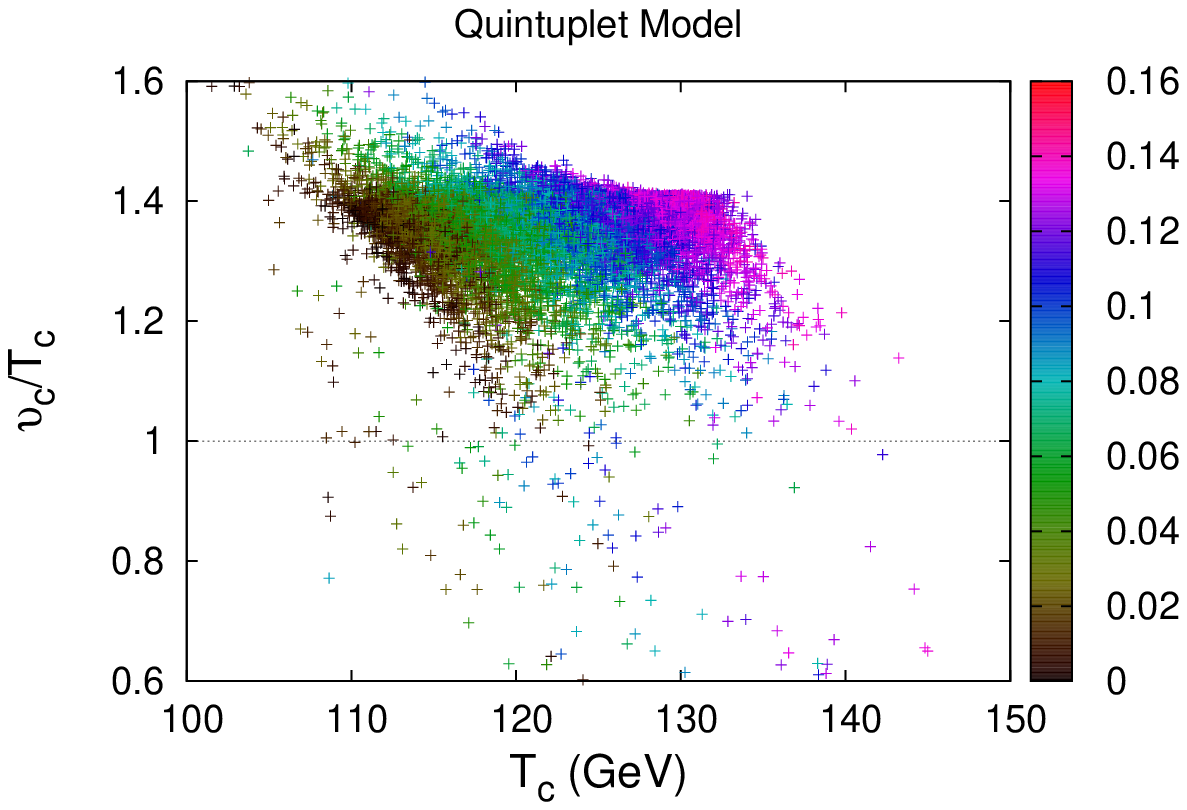}
\end{center}
\caption{\textit{The ratio $\upsilon(T_c)/T_c$ versus the critical
temperature for the (left) triplet and (right) quintuplet models.
The palette shows the Higgs quartic coupling $\lambda$.}}
\label{vctc}
\end{figure}

In Fig.~\ref{vctc}, we plot the ratio $\upsilon(T_c)/T_c$ verses the
critical temperature $T_c$, using the 20,000 benchmark points for the
triplet and quintuplet models. The figure shows that the EWPT is strongly
first-order for a majority of the benchmark sets, with the ratio $%
\upsilon(T_c)/T_c$ predominantly taking values between 1.2 and 1.5 in both
the triplet and quintuplet models. The transition temperature is a bit
larger than the typical SM value $\sim100$~\textrm{GeV}, and can be as large
as $150$~\textrm{GeV} and $140$~\textrm{GeV} for the triplet and quintuplet
models, respectively. One can read from the palettes in Fig.~\ref{vctc}
that, for fixed Higgs quartic coupling values, the ratio $\upsilon(T_c)/T_c$
is inversely proportional to the critical temperature.

Inspecting Fig.~\ref{vctc}, one is lead to the conclusion that the increased
EWPT strength is not only a consequence of a small Higgs quartic coupling $%
\lambda $, but can also be due to the transition dynamics; the existence of
heavy scalars makes the Higgs VEV slowly decaying with respect to the
temperature. Consequently the evolving (increasing or decreasing) effective
potential at the wrong vacuum makes the transition occurring at the
mentioned temperature values, therefore giving a large ratio $\upsilon
(T_{c})/T_{c}$.

\begin{figure}[h]
\begin{center}
\includegraphics[width=0.6\textwidth]{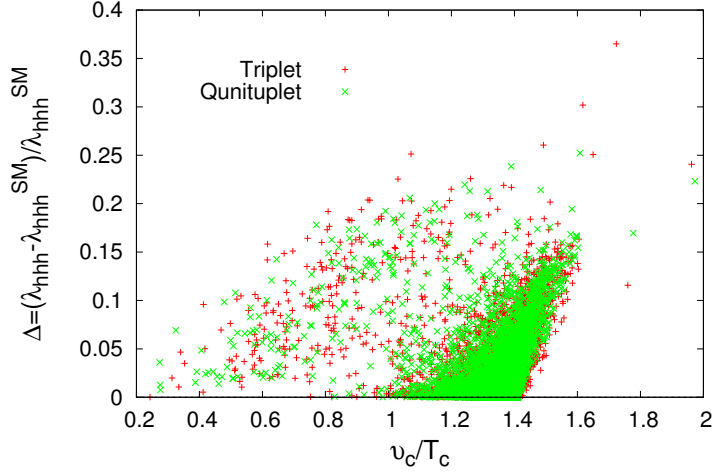}
\end{center}
\caption{\textit{The relative enhancement in the triple Higgs
coupling $\Delta$ versus the ratio $\upsilon(T_{c})/T_{c}$ for the
benchmark points used previously.}} \label{Dvc}
\end{figure}

An interesting issue, discussed in the literature \cite{Kanemura:2004ch}, is
a possible correlation between the EWPT strength and the relative
enhancement in the triple Higgs coupling $\Delta $. In Ref.~\cite%
{Ahriche:2015mea} it was shown that such a correlation is not clear for a
model with extra charged and neutral scalars from two inert doublets. In
Fig.~\ref{Dvc}, we plot the relative enhancement of the triple Higgs
coupling, $\Delta $, versus the ratio $\upsilon (T_{c})/T_{c}$, for the
20,000 benchmarks for the triplet and quintuplet models. From Figure \ref%
{Dvc}, it is not clear whether a correlation between the relative
enhancement in the triple Higgs coupling and the EWPT strength exists. This
issue deserves a detailed and model-independent investigation.

\section{The Higgs decay channels $h\rightarrow\gamma\gamma$
and $h\rightarrow\gamma Z$\label{sec:higgs_decays}}

In July 2012, the ATLAS~\cite{Aad:2012tfa} and CMS~\cite{Chatrchyan:2012xdj}
collaborations announced the observation of a scalar particle with mass $%
\simeq 125$~\textrm{GeV}, with roughly $5\sigma $ confidence level.
Subsequently this value was updated to $m_{h}=125.09\pm 0.21$~\textrm{GeV}
\cite{Aad:2015zhl}. An important question is whether or not this really is
the SM Higgs or an alternative Higgs-like state with different properties.
Indeed, a fit to the data, performed by both the ATLAS and CMS
collaborations, seems to (almost) show agreement with the SM, with the
reported values being $1.17\mp 0.27$~\cite{Aad:2014eha}\ and $1.13\pm 0.24$~%
\cite{CMS}, respectively.

Defining $R_{\gamma \gamma }$ as the branching ratio of $h\rightarrow \gamma
\gamma $ scaled by the SM value, we find that the present models give
\begin{equation}
R_{\gamma \gamma }=\frac{\mathcal{B}(h\rightarrow \gamma \gamma )}{\mathcal{B%
}^{SM}(h\rightarrow \gamma \gamma )}=\left\vert 1+\frac{\upsilon ^{2}}{2}%
\frac{\sum_{i}\frac{\vartheta _{i}}{m_{X_{i}}^{2}}A_{0}^{\gamma
\gamma }\left( \tau _{i}\right) }{A_{1}^{\gamma \gamma }\left( \tau
_{W}\right) +N_{c}Q_{t}^{2}A_{1/2}^{\gamma \gamma }\left( \tau
_{t}\right) }\right\vert ^{2}, \label{Ryy}
\end{equation}%
where $i$ stands for all charged scalars $X_{i}$, $\tau
_{i}=m_{h}^{2}/4m_{X_{i}}^{2}$, with $m_{X_{i}}$ being the mass of the
charged particle $X_{i}$ running inside the loop, $N_{c}=3$ is the color
number, and $Q_{t}$ is the top quark electric charge in units of $\left\vert
e\right\vert $. The parameters $\vartheta _{i}$ are given for triplet and
quintuplet members in Table \ref{T1}. In the above, the loop amplitudes $%
A_{k}^{\gamma \gamma }$ for spin $0$, spin $1/2$ and spin $1$ particles are
given by \cite{Djouadi:2005gi}
\begin{align}
A_{0}^{\gamma \gamma }\left( x\right) & =-x^{-2}\left[ x-f\left( x\right) %
\right] , \notag \label{A} \\
A_{1/2}^{\gamma \gamma }\left( x\right) & =2x^{-2}\left[ x+\left( x-1\right)
f\left( x\right) \right] , \notag \\
A_{1}^{\gamma \gamma }\left( x\right) & =-x^{-2}\left[ 2x^{2}+3x+3\left(
2x-1\right) f\left( x\right) \right] , \\
f\left( x\right) & =\left\{
\begin{array}{ccc}
\arcsin ^{2}\left( \sqrt{x}\right) & & x\leq 1 \\
-\frac{1}{4}\left[ \log \frac{1+\sqrt{1-x^{-1}}}{1-\sqrt{1-x^{-1}}}-i\pi %
\right] ^{2} & & x>1.%
\end{array}%
\right.
\end{align}

Another important Higgs decay channel that can be modified by the presence
of extra charged scalars is $h\rightarrow \gamma Z$. This channel is
similarly parameterized as
\begin{equation}
R_{\gamma Z}=\frac{\mathcal{B}(h\rightarrow \gamma Z)}{\mathcal{B}%
^{SM}(h\rightarrow \gamma Z)}=\left\vert 1+s_{\mathrm{w}}c_{\mathrm{w}%
}\upsilon ^{2}\frac{\sum_{i}\frac{\varkappa _{i}}{m_{X_{i}}^{2}}%
A_{0}^{\gamma Z}\left( \tau _{i},\zeta _{i}\right) }{c_{\mathrm{w}%
}^{2}A_{1}^{\gamma Z}\left( \tau _{W},\zeta _{W}\right) +2\left( 1-8s_{%
\mathrm{w}}^{2}/3\right) A_{1/2}^{\gamma Z}\left( \tau _{t},\zeta
_{t}\right) }\right\vert ^{2}, \label{RyZ}
\end{equation}%
where $\zeta _{i}=m_{Z}^{2}/4m_{X_{i}}^{2}$, and the functions $%
A_{k}^{\gamma Z}$ are given by \cite{Djouadi:2005gi}%
\begin{align}
A_{0}^{\gamma Z}\left( x,y\right) & =I_{1}\left( x,y\right) , \notag \\
A_{1/2}^{\gamma Z}\left( x,y\right) & =I_{1}\left( x,y\right) -I_{2}\left(
x,y\right) , \notag \\
A_{1}^{\gamma Z}\left( x,y\right) & =\left[ \left( 1+2x\right) s_{\mathrm{w}%
}^{2}/c_{\mathrm{w}}^{2}-\left( 5+2x\right) \right] I_{1}\left( x,y\right)
+4\left( 3-s_{\mathrm{w}}^{2}/c_{\mathrm{w}}^{2}\right) I_{2}\left(
x,y\right) , \\
I_{1}\left( x,y\right) & =-\tfrac{1}{2\left( x-y\right) }+\tfrac{f\left(
x\right) -f\left( y\right) }{2\left( x-y\right) ^{2}}+\tfrac{y\left[ g\left(
x\right) -g\left( y\right) \right] }{\left( x-y\right) ^{2}},~I_{2}\left(
x,y\right) =\tfrac{f\left( x\right) -f\left( y\right) }{2\left( x-y\right) },
\notag \\
g\left( x\right) & =\left\{
\begin{array}{ccc}
\sqrt{x^{-1}-1}\arcsin \left( \sqrt{x}\right) & & x\leq 1 \\
\frac{\sqrt{1-x^{-1}}}{2}\left[ \log \frac{1+\sqrt{1-x^{-1}}}{1-\sqrt{%
1-x^{-1}}}-i\pi \right] & & x>1.%
\end{array}%
\right.
\end{align}%
The parameters $\varkappa _{i}$ are shown in Table~\ref{T1} for both the
triplet and quintuplet models.

\begin{table}[t]
\begin{center}
$%
\begin{tabular}{|c|c|c|c|}
\hline
Model & Charged fields & $\vartheta_{i}$ & $\varkappa_{i}$ \\ \hline\hline
& $S^{+}$ & $\lambda_{SH}$ & $\frac{s_{w}}{c_{w}}\lambda_{SH}$ \\ \cline{2-4}
Triplet & $T^{+}$ & $\frac{2\lambda_{HT}-\bar{\lambda}_{HT}}{2}$ & $\frac{%
s_{w}}{c_{w}}\frac{2\lambda_{HT}-\bar{\lambda}_{HT}}{2}$ \\ \cline{2-4}
& $T^{++}$ & $4\lambda_{HT}$ & $-\frac{2-4s_{w}^{2}}{s_{w}c_{w}}\lambda_{HT}$
\\ \hline\hline
& $S^{+}$ & $\lambda_{SH}$ & $\frac{s_{w}}{c_{w}}\lambda_{SH}$ \\ \cline{2-4}
& $T^{-}$ & $\lambda_{HT1}+\lambda_{HT2}$ & $\frac{2+s_{w}^{2}}{s_{w}c_{w}}%
\left( \lambda_{HT1}+\lambda_{HT2}\right) $ \\ \cline{2-4}
Quintuplet & $T^{+}$ & $\frac{2\lambda_{HT1}+\lambda_{HT2}}{2}$ & $\frac {%
s_{w}}{c_{w}}\frac{2\lambda_{HT1}+\lambda_{HT2}}{2}$ \\ \cline{2-4}
& $T^{++}$ & $4\lambda_{HT1}+\lambda_{HT2}$ & $-\frac{1-2s_{w}^{2}}{%
s_{w}c_{w}}\frac{4\lambda_{HT1}+\lambda_{HT2}}{2}$ \\ \cline{2-4}
& $T^{+++}$ & $9\lambda_{HT1}$ & $-\frac{6-9s_{w}^{2}}{s_{w}c_{w}}%
\lambda_{HT1}$ \\ \hline
\end{tabular}
\ $%
\end{center}
\caption{The parameters $\vartheta_{i}$ and $\varkappa_{i}$, which
are relevant for the Higgs decay channels $h\rightarrow\gamma%
\gamma$\ and $h\rightarrow \gamma Z$.} \label{T1}
\end{table}

The deviation of the channels $h\rightarrow \gamma \gamma ,\gamma Z$ from
their SM values is sensitive to the mass of the scalars and the strength
with which they couple to the Higgs doublet, i.e.~on the parameters $%
m_{S}^{2}$, $m_{+}^{2}$, $m_{++}^{2}$, $\lambda _{SH}$, $\lambda _{HT}$ and $%
\bar{\lambda}_{HT}$ for the triplet model, and on $m_{S}^{2}$, $m_{+}^{2}$, $%
m_{++}^{2}$, $m_{+++}^{2}$, $\lambda _{SH}$ and $\lambda _{HT1,2}$ for the
quintuplet model. Depending on the relative sign of the couplings to the
Higgs doublet, the new contributions can strengthen or weaken the deviation
of $\mathcal{B}(h\rightarrow \gamma \gamma )$ from its SM value. In Fig.~\ref%
{Hr}, we present $R_{\gamma \gamma }$ versus $R_{\gamma Z}$ for the
considered 20,000 sets of benchmark parameters for both the triplet and
quintuplet models.

\begin{figure}[h]
\begin{centering}
\includegraphics[width=0.6\textwidth]{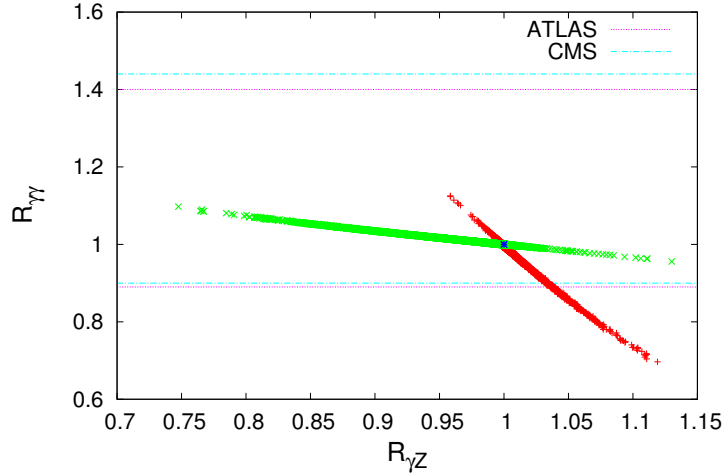}
\par\end{centering}
\caption{\textit{The modified Higgs decay rates $B(h\rightarrow\gamma%
\gamma)$ vs $B(h\rightarrow\gamma Z)$, scaled by their SM values,
due to the extra charged scalars, for 20,000 randomly chosen sets of
benchmark parameters for the triplet (red) and quintuplet (green)
models. The intervals between the magenta (green) lines represent
the ATLAS (CMS) recent measurements on the $h\rightarrow
\gamma\gamma$ channel, and the blue point represents the SM.}}
\label{Hr}
\end{figure}

We remark that some benchmarks in the triplet model are already excluded by
the recent measurements of ATLAS \cite{Aad:2015zhl} and CMS \cite{CMS},
while more precise measurements are required to probe benchmarks for the
quintuplet model. In contrast to other models with extra singlets~\cite%
{Ahriche:2013zwa} and doublets \cite{Ahriche:2015mea}, the decay channel $%
h\rightarrow \gamma Z$\ can be significantly modified, with respect to the
SM value, particularly in the quintuplet model. This can be understood from
the large $\varkappa _{i}$ coupling values for the scalar multiplet members
in Table~\ref{T1}.

\section{Conclusion\label{sec:conc}}

Models of radiative neutrino mass with DM candidates can explain some of the
short-comings of the SM while generating observable experimental signals. In
this work, we performed a detailed study of the scalar-sector phenomenology
for a pair of three-loop neutrino mass models with DM candidates. The
models, referred to as the triplet~\cite{Ahriche:2014cda} and quintuplet~%
\cite{Ahriche:2014oda} models, generate neutrino mass via a diagram with the
same topology as the KNT model. We investigated the effect of the extra
scalars on the Higgs mass, the triple Higgs coupling, and the Higgs decay
channels $h\rightarrow\gamma \gamma, \gamma Z$. We also studied the strength
of the electroweak phase transition. In both models, it was shown that the
beyond-SM multiplets can modify the triple Higgs coupling and the Higgs
decay channels away from their SM values. The electroweak phase transition
was found to be strongly first-order in significant regions of parameter
space. Measurements of the Higgs decay channels already exclude some regions
of parameter space for the triplet model, and future improvements will
further explore the parameter space for both models.

\section*{Acknowledgments\label{sec:ackn}}

AA thanks the ICTP for the hospitality during the last stage of this work.
AA is supported by the Algerian Ministry of Higher Education and Scientific
Research under the CNEPRU Project No D01720130042. KM is supported by the
Australian Research Council.

\appendix

\section{Field Dependent Masses\label{app:mass}}

The charged scalar and SM field-dependent masses are given by:
\begin{align}
m_{\chi }^{2}& =-\mu ^{2}+\lambda h^{2}+\Pi _{H},~m_{h}^{2}=-\mu
^{2}+3\lambda h^{2}+\Pi _{H},~m_{t}^{2}=\frac{y_{t}^{2}}{2}%
h^{2},~m_{W}^{2}=m_{W_{3}W_{3}}^{2}=\frac{g_{2}^{2}}{4}h^{2}+\Pi _{W},~
\notag \\
m_{BB}^{2}& =\frac{g_{1}^{2}}{4}h^{2}+\Pi _{B},~m_{W_{3}B}^{2}=\frac{%
g_{2}g_{1}}{4}h^{2},~m_{S}^{2}=\mu _{S}^{2}+\frac{\lambda _{SH}}{2}h^{2}+\Pi
_{S},
\end{align}%
with%
\begin{align}
\Pi _{H}& =\left( 12\lambda +9g^{2}+3g^{\prime 2}+3y_{t}^{2}+2\lambda
_{SH}\right) \frac{T^{2}}{12},~\Pi _{S}=\left( 4\lambda _{SH}+4\lambda _{{S}%
}+3g^{\prime 2}\right) \frac{T^{2}}{12}, \notag \\
\Pi _{W}^{L}& =\Pi _{W_{3}}^{L}=\frac{11}{6}g^{2}T^{2},~\Pi _{B}^{L}=\frac{27%
}{16}g^{\prime 2}T^{2},~\Pi _{W}^{T}=\Pi _{B}^{T}=0.
\end{align}%
Here, we ignored the triplet and quintuplet contributions since they
decouple from the thermal plasma due to their large masses, relative to the
relevant typical temperature of $\mathcal{O}(100~\mathrm{GeV})$.

The triplet members field dependant masses are given by:%
\begin{eqnarray}
m_{0}^{2} &=&\mu _{T}^{2}+\frac{\lambda _{HT}-\bar{\lambda}_{HT}}{2}%
h^{2}+\Pi _{T},~m_{+}^{2}=\mu _{T}^{2}+\frac{2\lambda _{HT}-\bar{\lambda}%
_{HT}}{4}h^{2}+\Pi _{T}, \notag \\
m_{++}^{2} &=&\mu _{T}^{2}+\frac{\lambda _{HT}}{2}h^{2}+\Pi _{T},~\Pi
_{T}=\left( 9g^{2}+3g^{\prime 2}+2\bar{\lambda}_{HT}+4\lambda _{HT}+2\bar{%
\lambda}_{ST}\right) \frac{T^{2}}{12},
\end{eqnarray}%
and the quintuplet members field dependant masses are given by:%
\begin{eqnarray}
m_{-}^{2} &=&\mu _{T}^{2}+\frac{\lambda _{HT1}+\lambda _{HT2}}{2}h^{2}+\Pi
_{T},~m_{0}^{2}=\mu _{T}^{2}+\frac{4\lambda _{HT1}+3\lambda _{HT2}}{8}%
h^{2}+\Pi _{T}, \notag \\
m_{+}^{2} &=&\mu _{T}^{2}+\frac{2\lambda _{HT1}+\lambda _{HT2}}{4}h^{2}+\Pi
_{T},~m_{++}^{2}=\mu _{T}^{2}+\frac{4\lambda _{HT1}+\lambda _{HT2}}{8}%
h^{2}+\Pi _{T}, \notag \\
~m_{+++}^{2} &=&\mu _{T}^{2}+\frac{\lambda _{HT1}}{2}h^{2}+\Pi _{T},~\Pi
_{T}=\left( 9g^{2}+3g^{\prime 2}+2\lambda _{ST}+4\lambda _{HT1}+2\lambda
_{HT2}\right) \frac{T^{2}}{12}.
\end{eqnarray}

\end{document}